\documentstyle[12pt]{article}

\newcommand{\bm}[1]{\mbox{\boldmath $#1$}}

\begin{document}
\begin{flushright}
DTP--MSU/97-04\\ 
February 4, 1997
\\  gr-qc/yymmxxx
\end{flushright}  
\vskip1.5cm
\begin{center}
{\LARGE\bf
Matrix Ernst potentials for EMDA with multiple vector fields}
\vskip1cm
{\bf
D.V. Gal'tsov}\footnote{Email: galtsov@grg.phys.msu.su} 
and {\bf S.A. Sharakin}
\footnote{Email:sharakin@grg1.phys.msu.su}
\\
\normalsize Department of Theoretical Physics, Moscow State University,\\ 
\normalsize Moscow 119899, {\bf Russia}\\ 
\vskip1cm

\end{center}
\begin{abstract}
We show that the Einstein--Maxwell--Dilaton--Axion system with multiple
vector fields (bosonic sector of the $D=4, N=4$ supergravity) restricted
to spacetimes possessing a non--null Killing vector field admits
a concise representation in terms of the Ernst--type matrix valued 
potentials. A constructive derivation of the SWIP solutions is given 
and a colliding waves counterpart of the DARN-NUT solution is obtained. 
$SU(m,m)$ chiral representation of the two--dimensionally reduced system 
is derived and the corresponding Kramer--Neugebauer--type map is presented. 
\vskip5mm
\noindent
PACS number(s): 97.60.Lf, 04.60.+n, 11.17.+y
\end{abstract}

Recently a variety of black hole solutions was found in the 
four--dimensional extended supergravities \cite{tudo} using 
either {\it ad hoc} ansatze or employing classical dualities. 
In the most extensively studied  $N=4$ theory it was shown 
that the corresponding three-dimensional reduction (with a non--null 
spacetime Killing symmetry assumed) may be concisely formulated in terms
of generalized Ernst potentials \cite{ad}. This suggests an alternative
interpretation of the classical $U$--duality as the `Ehlers'
symmetry and opens a way to apply 
powerful general relativity techniques to construct exact classical 
solutions. For the Einstein--Maxwell--Dilaton--Axion (EMDA) theory with 
one vector field a particularly simple matrix Ernst potential was found
in terms of $2\times 2$ symmetric complex  matrices \cite{gk}. 
This representation, however, is due to existence of an exceptional 
local isomorphism $SO(2,3)\sim Sp(4,R)$, relevant to the one-vector 
EMDA $U$--duality $SO(2,3)$ \cite{g}, which is not 
extendible to the realistic case of multiple vector fields. Here we show 
that in the case of two vector fields ($p=2$) another exceptional local 
isomorphism $SO(2,4)\sim SU(2,2)$ gives rise to an even more economical 
representation of the $8$--dimensional TS in terms of the $2\times 2$ 
complex {\em non--symmetric} 
matrices (reducing to symmetric ones for $p=1$). For arbitrary $p$ 
a matrix Ernst potential can be constructed using the Clifford algerbras 
corresponding to the compact subgroup $SO(p+1)$ of the three--dimensional 
$T$--duality group $SO(1,p+1)$. 
This leads to the pseudounitary embedding of the $U$--duality group 
$SO(2,2+p)$ into $SU(m,m)$ where $m=2^k,\, k=[(p+1)/2]$. 
In terms of the matrix Ernst potential $U$--duality looks like
a matrix--valued `Ehlers' $SL(2,R)$ symmetry \cite{eh}. Further
two--dimensional reduction of the theory (with the rank--two Abelian
spacetime isometry group assumed) leads to the $SU(m,m)$ chiral
representation in the $\sigma$--model variables, or to its `Matzner--Misner' 
counterpart obtainable via the Kramer--Neugebauer--type map.

We start with the four--dimensional action
\begin{equation}
S=\int \left\{-R+2\left|\partial z (z-{\bar z})^{-1}\right|^2 +
\left(iz{\cal F}_{\mu\nu}^{n}{\cal F}^{n \mu\nu}+c.c\right)\right\}
\sqrt{-g}d^4x,
\end{equation}
where ${\cal F}^n=(F^n+i{\tilde F}^n)/2,\;
{\tilde F}^{n \mu\nu}=\frac{1}{2}E^{\mu\nu\lambda\tau}F^n_{\lambda\tau}, \,
n=1,...,p$ (summation over repeated $n$ is understood elsewhere), 
$z=\kappa+ie^{-2\phi}$, and the metric signature is $+---$. For $p=6$ this
action describes the bosonic sector of $N=4, D=4$ supergravity. 
It is invariant under the $SO(p)$ rotation of the vector fields, which is 
an analogue of the $T$--duality of dimensionally reduced theories \cite{gpr}.
The equations of motion and Bianchi identities (but not the action) are also
invariant under the $S$--duality transformations
$$
z \rightarrow \frac{az+b}{cz+d},\quad ad-bc=1,
$$
\begin{equation}
F^n \rightarrow (c\kappa+d) F^n
+ c {\rm e}^{-2\phi} {\tilde F}^n.
\end{equation}

Consider three--dimensional reduction of the theory 
assuming either timelike ($\lambda=1$), or spacelike ($\lambda=-1$)
(in an essential region of spacetime) Killing symmetry. Then the 
four--dimensional line element may be written as
\begin{equation}
ds^2=\lambda f(dy-\omega_idx^i)^2-
\frac{\lambda}{f}h_{ij}dx^idx^j,
\end{equation}
where the three--space metric $h_{ij}\;(i, j=1, 2, 3)$, the 
one--form $\omega_i $ and the conformal factor $f$
depend on the three--space coordinates $x^i$ only. It is assumed
that $y=t,\, h_{ij}$ is spacelike for $\lambda=1$, and $h_{ij}$ is of the 
signature $+--$ for $\lambda=-1$.

One can express vector fields
through the quantities $v^n,\,u^n$ which have the  meaning of the electric 
and magnetic scalar potentials for $\lambda=1$:
\begin{equation}
F^n_{iy}=\frac{1}{\sqrt{2}}\partial_iv^n,
\end{equation}
\begin{equation}
2{\rm Im}\left( z{\cal F}^{n ij}\right)
=\frac{f}{\sqrt{2h}}\epsilon^{ijk}\partial_ku^n,\quad 
h\equiv \det h_{ij}. 
\end{equation}
In three dimensions the `$T$--duality' group is enlarged to $SO(1,p+1)$, 
while the $S$--duality becomes the symmetry of the action. Moreover,
both these groups are unified within a larger `$U$--duality'
group $SO(2,p+2)$ \cite{gk,udu,ht}.
This can be easily shown by constructing the K\"ahler
metric of the target manifold of the resulting $\sigma$--model. To find
such a representation one has to introduce a twist potential $\chi$ via
\begin{equation}
d\chi=u^n d v^n -v^n d u^n -\lambda f^2 *d \omega,
\end{equation}
and to derive equations for $\chi, u^n$ in addition to those
for $f, \kappa, \phi, v^n$. The full set of equations will be that of 
the three--dimensional gravity coupled non--linear $\sigma$--model
possessing the $4+2p$ dimensional target space
$SO(2,2+p)/\left(SO(2)\times SO(p,2)\right)$ for $\lambda=1$, respectively
$SO(2,2+p)/\left(SO(2)\times SO(p+2)\right)$ for $\lambda=-1$. 
In the latter case the corresponding matrices are symmetric, what is
a desirable property for an application of the inverse scattering
transfrom technique. Since the transition from $\lambda=1$
to $\lambda=-1$ in (3) is merely an analytic continuation, symmetric
matrices may be used in the $\lambda=1$ case as well (a realization 
of the non--compact coset by symmetric matrices may be
achieved via the left multiplication by some constant matrix).

The target manifold can be parametrized by complex coordinates 
$z^\alpha,\, \alpha=0,1,...,p+1$
which have the following meaning. The components $\alpha=n=1,...,p$
are complex potentials for vector fields
\begin{equation}
z^n = u^n-z v^n\equiv \Phi^n, \quad n=1,...,p,
\end{equation}
while the $\alpha=0, p+1$ components are linear combinations of 
the complex axidilaton and the Ernst potential $E=i\lambda f-\chi
+v^n\Phi^n$:
\begin{equation}
z^0 = (E-z)/2,\quad z^{p+1}=(E+z)/2.
\end{equation}
The TS metric is generated by the K\"ahler potential \cite{ad}
\begin{equation}
G_{\alpha {\bar \beta}}=\partial_\alpha \partial_{\bar\beta}
K(z^\alpha, {\bar z}^\beta),  
\end{equation}
\begin{equation}
K=-\ln V,\quad V=\lambda\eta_{\alpha\beta}{\rm Im} z^\alpha {\rm Im} z^\beta
=f{\rm e}^{-2\phi},
\end{equation}
where the $T$--duality $SO(1,p+1)$ metric is introduced
$\eta_{\alpha\beta}=diag (-1,1,...,1),\,\alpha, \beta=0,1,...,p+1$.

For $p=1$ the matrix Ernst potential incorporating linearly all 
K\"ahler variables reads \cite{gk,udu}
\begin{equation}
{\cal E}=\pmatrix{
E & \Phi \cr
\Phi & -z}.
\end{equation}
This is a symmetric complex matrix which splits into 
hermitean and  antihermitean parts 
\begin{equation}
{\cal E} = {\cal Q}+i{\cal P},\quad {\cal P}^{\dagger}={\cal P},\;\; 
{\cal Q}^{\dagger}={\cal Q},
\end{equation}
with {\it real} ${\cal Q,\, P}$ and generates a symmetric $Sp(4,R)$ matrix
\begin{equation}
{\cal M}=\left(\begin{array}{crc}
{\cal P}^{-1}&{\cal P}^{-1}{\cal Q}\\
{\cal Q}{\cal P}^{-1}&{\cal P}+{\cal Q}{\cal P}^{-1}{\cal Q}\\
\end{array}\right),
\end{equation}
satisfying
\begin{equation}
{\cal M}^{\dagger} J {\cal M} = J,\quad
J=\pmatrix{
O & I_2 \cr
-I_2 & O}.
\end{equation}
It can be checked that the TS metric is 
\begin{equation}
dl^2=-2{\rm Tr}\left\{d{\cal E} \left({\cal E}^{\dagger}-
{\cal E}\right)^{-1}
d{\cal E}^{\dagger}\left( {\cal E}^{\dagger}-{\cal E}\right)^{-1}\right\}.
\end{equation}
K\"ahler potentials act as scalar sources in the three--dimensional
Einstein's equations
\begin{equation}
{\cal R}_{ij}=- 2{\rm Tr}\left\{\left({\cal E}^{\dagger}-
{\cal E}\right)^{-1}
\left(\partial_{(i}{\cal E}\right) \left({\cal E}^{\dagger}-
{\cal E}\right)^{-1}
\partial_{j)}{\cal E}^{\dagger}\right\}.
\end{equation}
Alternatively, in terms of ${\cal M}$, the TS metric reads
\begin{equation}
dl^2=-\frac{1}{4} {\rm Tr} \{d{\cal M}d{\cal M}^{-1}\},
\end{equation}
while the Einstein's equations for $h_{ij}$ are
\begin{equation}
{\cal R}_{ij}=-\frac{1}{4} {\rm Tr} \{
\left(\partial_{(i} {\cal M}\right)
\partial_{j)}{\cal M}^{-1} \}.
\end{equation}

Here we are looking for a generalization of this representation
to higher $p$. It turns out that this can be achieved not in terms
of higher rank symplectic groups, but rather in terms of pseudounitary
imbeddings. Consider first the case $p=2$. Then the  global symmetry 
of the TS ($U$-duality) is the four--dimensional conformal group 
$SO(2,4)\sim SU(2,2)$. The latter group, realised 
by $(4\times 4)$ complex matrices, can be conveniently presented using 
the Dirac basis $\sigma_{\mu\nu}=\rho_{\mu}\otimes\sigma_{\nu}$,
where $\rho_{\mu},\,\sigma_{\nu}$ are two sets of Pauli matrices 
(with the unit matrix for $\mu,\nu=0$) \cite{cg}. Any element
${\cal U}\in SU(2,2)$ satisfies
${\cal U}^{\dagger} \sigma_{30} {\cal U} = \sigma_{30}$.
To get contact with $p=1$ one has to perform the unitary transformation
\begin{equation}
{\cal M}=V^{\dagger} {\cal U} V, \quad V = (\sigma_{00} - i\sigma_{10})/
\sqrt{2},
\end{equation}
so that ${\cal M}$ should obey (14) (in the
context of unitary groups it is more natural to multiply
$J$ by $i$, {\em i.e.} to take $J=\sigma_{20}$). Then the expression 
(15) for the TS line element remains valid (up to a numerical factor) for 
the following $p=2$ matrix Ernst potential:
\begin{equation}
{\cal E}=\pmatrix{
E & \Phi_1-i\Phi_2 \cr
\Phi_1+i\Phi_2 & -z}.
\end{equation}
With the same block parametrization (13)  
the formulas (17,18) also hold up to a normalization. Note that
now hermitean ${\cal P, Q}$ are not real.

The essential feature of the matrix Ernst representation is that
it provides the {\em matrix--valued} generalization of the Ehlers
group of the vacuum general relativity \cite{ad}. This gives an 
alternative view on the $U$--duality in three--dimensional 
supergravities. For $p=2$ the 15--parametric 
`Ehlers' group consists of the four--parametric gauge,
\begin{equation}
{\cal E }\rightarrow {\cal E }+{\cal G }, 
\qquad {\cal G }=
\pmatrix{
g & m^1-im^2 \cr
m^1+im^2 & b},
\end{equation}
($g,\,b$ are twist and axion shift parameters, $m^n$  is a magnetic gauge),
the four--parametric `proper Ehlers' (including the `Ehlers'--like
$S$--duality component),
\begin{equation}
{\cal E }^{-1}\rightarrow {\cal E }^{-1}+{\cal H }, 
\qquad {\cal H }=
\pmatrix{
c_E & h_m^1-ih_m^2 \cr
h_m^1+ih_m^2 & c},
\end{equation}
and the seven--parametric `scale' transformation:
\begin{equation}
{\cal E }\rightarrow {\cal S}^{\dagger} {\cal E }{\cal S},
\qquad {\cal S }=
\pmatrix{
e^{s+i\alpha} & h_e^1-ih_e^2 \cr
-e^1+ie^2 & a e^{-i\alpha}}.
\end{equation}
Note, that the Harrison transformations of this theory \cite{ad,gl}
(parametrized by $h_e^n,\, h_m^n,\, n=1,2$) enter partly into the
`Ehlers' and partly into the `scale' subgroups. In the latter
the parameter $\alpha$ represents the $SO(p)\, (p=2)$ rotations 
(the four--dimensional `$T$--duality').

To get the desired generalization to arbitrary $p$ the following 
observation is appropriate. The structure of the matrix Ernst potential
for $p=2$ may be viewed as an expansion over the Clifford algebra
corresponding to the $SO(p+1)$ subgroup of the three--dimensional
$T$--duality group:
\begin{equation}
\left\{\gamma_a,\, \gamma_b\right\}=2\delta_{ab}I_m,
\end{equation}
where $a,b=1,....,p+1,\, m=2^k,\, k=[(p+1)/2]$. For $p=2$ the Clifford
algebra is realized by the Pauli matrices $\sigma_a$, and clearly
\begin{equation}
{\cal E}=z^0 I_2+z^a\sigma_a.
\end{equation}
For arbitrary $p$ one has merely to replace $\sigma_a$ by 
{\em hermitean} $\gamma_a$:
\begin{equation}
{\cal E}=z^0 I_k+z^a\gamma_a.
\end{equation}
The dimensionality of this representation follows the usual rule
valid for gamma--matrices in arbitrary dimensions:
for even $p=2k$ gamma-matrices are $2^k\times 2^k$, while for $p=2k+1$
the rank is the same as for $p=2(k+1)$. The only modification to be made
in (15-18) is a numerical factor due to the trace of the unit matrix:
\begin{equation}
dl^2=
\frac{1}{m}{\rm Tr}\left\{d{\cal E}{\cal P}^{-1}  
d{\cal E}^{\dagger}{\cal P}^{-1}\right\} =
-\frac{1}{2m} {\rm Tr} \left\{d{\cal M}d{\cal M}^{-1}\right\},
\end{equation}
\begin{equation}
{\cal R}_{ij}=\frac{1}{m}{\rm Tr}\left\{
\left(\partial_{(i} {\cal E}\right){\cal P}^{-1} 
\left(\partial_{j)}{\cal E}^{\dagger}\right){\cal P}^{-1}\right\}=
-\frac{1}{2m} {\rm Tr} \left\{
\left(\partial_{(i} {\cal M}\right)
\partial_{j)}{\cal M}^{-1} \right\}.
\end{equation}
The corresponding expansions of ${\cal Q}, {\cal P}$ are given by 
(26) again with the real and imaginary parts of $z^{\alpha}$ respectively.
Matrices ${\cal M}$ are hermitean
by construction (for both $\lambda=\pm 1$) and belong to $SU(m,m)$. The
complex matrices ${\cal E}$ are `filled densely' only for $p=2$, in which
case the number of complex potentials coincides with the number 
of matrix elements (four). For $p>2$ one has $m^2 > p+2$.
This reflects the fact that 
the local isomorphism between  $SO(2,p+2)$ and non--compact unitary groups 
holds uniquely for $p=2$, for higher $p$ we deal only with 
{\em embeddings} into $SU(m,m)$.

Consider now the case $p=2$ in more detail. The algebra $su(2,2)$ is
formed by the complex traceless $4\times 4$ matrices $X$ subject to the
condition
\begin{equation}
X^{\dagger}\sigma_{20}+\sigma_{20}X^{\dagger}=0.
\end{equation}
It consists of 8 hermitean $\sigma_{10}$, $\sigma_{30}$, $\sigma_{11}$, 
$\sigma_{31}$, $\sigma_{12}$, $\sigma_{32}$, $\sigma_{13}$, $\sigma_{33}$, 
and 7 antihermitean $i(\sigma_{01}$, $\sigma_{02}$, $\sigma_{03}$,
$\sigma_{21}$, $\sigma_{22}$, $\sigma_{23}$, $\sigma_{20})$ generators.
Its subsequent decomposition will be performed in relation
to the geodesic ansatz for the matrix ${\cal M}$:
\begin{equation}
{\cal M}= A e^{B\sigma}.
\end{equation}
(More about geodesic technique with a detailed discussion of the $p=1$ 
theory see in \cite{cg}). In (30) $\sigma$ is a harmonic function on the
three--space, normalized to zero in some (`empty') region of the spacetime 
(where ${\cal M}= A$), and $B$ is an element of 
$su(2,2)$. We will be interested in two types of solutions: stationary 
asymptotically flat (SAF) configurations (elliptic case, $\lambda=1$),
and colliding plane waves (CPW) (hyperbolic case, $\lambda=-1$). For SAF
$A=\sigma_{03}$, while for CPW $A=-\sigma_{00}$ (this is equivalent 
to say that in the `empty' region $f=1, \chi=\phi=\kappa=v^n=u^n=0$). 
In both cases it is convenient to use
a representation with the hermitean ${\cal M}\in SU(2,2)$, therefore 
the matrix $B$ has to satisfy the following conditions
\begin{equation}
AB=B^{\dagger} A, \quad BK+KB=0, 
\end{equation}
where $ K=\sigma_{23}\;{\rm for \; SAF, and\;}
K=-\sigma_{20} \; {\rm for \;CPW}$. 
In the SAF case the elements of $su(2,2)$ satisfying these conditions
form two sets of $SO(2,2)$ Clifford algebras
\begin{equation}
\Gamma^1_{\mu}=\{i\sigma_{21},i\sigma_{22},-\sigma_{10},\sigma_{30}\}, \quad
\Gamma^2_{\mu}=\{i\sigma_{02},-i\sigma_{01},\sigma_{33},\sigma_{13},\},
\end{equation}
obeying
\begin{equation}
\{\Gamma^1_{\mu},\Gamma^1_{\nu}\}=\{\Gamma^2_{\mu},\Gamma^2_{\nu}\}=
2\eta_{\mu\nu}I
\end{equation} 
with $\eta_{\mu\nu}={\rm diag} (-1,-1,1,1)$.
The remaining generators span  the $so(2,2)\times so(2)={\cal H'}$ subalgebra
consisting of $$M_{12}=i\sigma_{03}/2,\;\; M_{13}=\sigma_{31}/2,\;\;
M_{14}=\sigma_{11}/2,\;\;$$ 
\begin{equation}
M_{23}=\sigma_{32}/2,\;\;
M_{24}=\sigma_{12}/2,\;\;  M_{34}=-i\sigma_{20}/2,
\end{equation}
$M_{\mu\nu}=-M_{\nu\mu}$, and $D=i\sigma_{23}$.

The commutation relations read
$$\left[D,M_{\mu\nu}\right]=0,\quad
\left[\Gamma^1_{\mu},\Gamma^1_{\nu}\right]=
\left[\Gamma^2_{\mu},\Gamma^2_{\nu}\right]=-4M_{\mu\nu},\quad
\left[\Gamma^1_{\mu},\Gamma^2_{\nu}\right]=2D\eta_{\mu\nu},$$
\begin{equation}
\left[D,\Gamma^1_{\mu}\right]=-2\Gamma^2_{\mu},\quad 
\left[M_{\mu\nu},\Gamma^1_{\lambda}\right]=
\eta_{\mu\lambda}\Gamma^1_{\nu}-\eta_{\nu\lambda}\Gamma^1_{\mu},
\end{equation}
$$\left[D,\Gamma^2_{\mu}\right]=2\Gamma^1_{\mu},\quad 
\left[M_{\mu\nu},\Gamma^2_{\lambda}\right]=
\eta_{\mu\lambda}\Gamma^2_{\nu}-\eta_{\nu\lambda}\Gamma^2_{\mu},$$
together with the standard commutators for $M_{\mu\nu}\in so(2,2)$. 
Also useful are the following anticommutators:
$$\left\{\Gamma^1_\mu, \Gamma^2_\nu\right\}=4 \tilde M_{\mu\nu},\quad
\left\{M_{\mu\nu},\Gamma^1_{\lambda}\right\}=
-i{\epsilon_{\mu\nu\lambda}}^\rho\Gamma^2_{\rho},\quad
\left\{M_{\mu\nu},\Gamma^2_{\lambda}\right\}
=i{\epsilon_{\mu\nu\lambda}}^\rho\Gamma^1_{\rho},$$
\begin{equation}
\left\{M_{\mu\nu},M_{\lambda\rho}\right\}=-\frac{1}{2} 
(\eta_{\mu\lambda}\eta_{\nu\rho}-\eta_{\mu\rho}\eta_{\nu\lambda})I -
\frac{i}{2}D\epsilon_{\mu\nu\lambda\rho}.
\end{equation}
where $\tilde M_{\mu\nu}=i{\epsilon_{\mu\nu}}^{\lambda\tau}
 M_{\lambda\tau}/2,\,\epsilon_{1234}=1$.

In the CPW case one deals with the Clifford algebras related to  the
compact subgroup $SO(4)$:
\begin{equation}
\Gamma^1_{\mu}=-\left\{\sigma_{11}, \sigma_{12}, 
\sigma_{13}, \sigma_{30}\right\}, \quad
\Gamma^2_{\mu}=\left\{\sigma_{31}, \sigma_{32}, 
\sigma_{33}, -\sigma_{10}\right\},
\end{equation}
while the remaining generators span the $so(4)\times so(2)={\cal H}$ 
(maximal compact) subalgebra of $su(2,2)$:
$$M_{12}=-i\sigma_{03}/2, \quad M_{13}=i\sigma_{02}/2, \quad
M_{14}=i\sigma_{21}/2,$$
\begin{equation}
M_{23}=-i\sigma_{01}/2,\quad M_{24}=i\sigma_{22}/2,\quad
M_{34}=i\sigma_{23}/2,\quad D=i\sigma_{20}.
\end{equation}
The commutators and anticommutators are the same, but now
$\eta_{\mu\nu}={\rm diag} (1,1,1,1)$. In both cases $\lambda=\pm 1$ 
we have:
\begin{equation}\label{B}
B=\bm\alpha {\bm\Gamma}^1+\bm\beta{\bm\Gamma}^2
\equiv\alpha^{\mu}\Gamma^1_{\mu}+
\beta^{\mu}\Gamma^2_{\mu}, 
\end{equation}
with  constant $SO(2,2),\,({\rm resp.}\, SO(4))$ vectors 
$\bm\alpha,\,\bm\beta$.
Similarly to \cite{cg}, one can show that 
\begin{eqnarray}
\label{B2}&& B^2=(\bm\alpha^2+\bm\beta^2)I+4(\bm\alpha\wedge\bm\beta)
\cdot\tilde {\bf M},\nonumber\\
&&B^3={\bm\alpha}'{\bm\Gamma}^1+{\bm\beta}'{\bm\Gamma}^2,\\
\label{B4}&& B^4=\left[(\bm\alpha^2+\bm\beta^2)^2+
4(\bm\alpha\wedge\bm\beta)^2\right]I
+8(\bm\alpha^2+\bm\beta^2)(
\bm\alpha\wedge\bm\beta)\cdot\tilde {\bf M},\nonumber
\end{eqnarray}
where
$\bm\alpha^2=\eta_{\mu\nu}\alpha^{\mu}\alpha^{\nu}$ etc., 
and
\begin{equation}
\bm\alpha'= 2\bm\beta\wedge(\bm\beta\wedge\bm\alpha)
+(\bm\alpha^2+\bm\beta^2)\bm\alpha,\quad
\bm\beta'= 2\bm\alpha\wedge(\bm\alpha\wedge\bm\beta)
+(\bm\alpha^2+\bm\beta^2)\bm\beta.
\end{equation}

Leaving the construction of the 
most general null geodesic solution to a separate publication, 
here we give the geodesic interpretation of the `SWIP' solutions
found recently \cite{bko}. 
They correspond to degenerate $B$. From (\ref{B}) one finds
\begin{equation}
{\rm det} B=(\bm\alpha^2-\bm\beta^2)^2+4(\bm\alpha \bm\beta)^2. 
\end{equation}
For SAF the standard definition \cite{cg} 
of the ADM mass $M$, the NUT parameter $N$, the dilaton and 
axion charges $D, A$ and electric/magnetic charges $Q^n, P^n$ (assuming 
$\sigma\rightarrow 1/r$ as $r\rightarrow \infty$) gives
\begin{equation}
\alpha^\mu=(\sqrt{2} P^1, \sqrt{2} P^2, A-N, M+D ),\quad
\beta^\mu=(-\sqrt{2} Q^1, -\sqrt{2} Q^2, M-D, A+N ).
\end{equation}
The degeneracy condition det$B=0$ implies $\bm\alpha^2=\bm\beta^2$ 
and $\bm\alpha \bm\beta=0$, what reduces to
\begin{equation}
D+iA=-\frac{\left(Q^n+iP^n\right)^2}{2(M+iN)}.
\end{equation}
This is a well--known relation for axion--dilaton black holes.

Extremal black holes can be identified with null geodesics. Since  
\begin{equation}
dl^2=\frac{1}{4}{\rm Tr}(B^2)(d\sigma)^2, \quad {\cal R}_{ij}=
\frac{1}{4}{\rm Tr} (B^2)\sigma_i\sigma_j,
\end{equation}
and in this case Tr($B^2)=0$, the three--space is 
Ricci--flat. According to (\ref{B2}), 
\begin{equation}
{\rm Tr}(B^2)=4(\bm\alpha^2+\bm\beta^2),
\end{equation}
so geodesics are null if $\bm\alpha^2+\bm\beta^2=0$
(what may be fulfiled with non--zero $\bm\alpha, \bm\beta$
in the $SO(2,2)$ case). For $p=1$ this condition entails 
$B^2=0$ ({\em i.e.} collinear  $\bm\alpha$ and $\bm\beta$ \cite{cg}), 
but for $p\ge 2$ it is not necessarily so. 

The $p=2$ TS admits four mutually orthogonal null vectors, and consequently 
four independent (real) harmonic functions may be incorporated into
the geodesic construction \cite{cg}. It is convenient to introduce 
$\sigma$--valued vectors $\bf a$ and $\bf b$ as linear combinations
$a^{\mu}=\alpha^{\mu}_{(A)}\sigma_{(A)}$ and  
$b^{\mu}=\beta^{\mu}_{(A)}\sigma_{(A)},\, A=1,...,4$, so that   
$B={\bf a}{\bm\Gamma}^1+{\bf b}{\bm\Gamma}^2$ (only four components
of $\sigma$--valued vectors are independent in view of the consistency
conditions \cite{cg}), then 
\begin{equation}
{\cal M}=A\left(I_4+{\bf a}{\bm\Gamma}^1+
{\bf b}{\bm\Gamma}^2+2({\bf a}\wedge {\bm b})\cdot\tilde {\bf M}\right).
\end{equation}
Comparing with (13) one finds
\begin{eqnarray}
&&f^{-1}=(1+a^4)(1+b^3)-a^3b^4,\quad
e^{2\phi}=f \left[(1+a^4)^2+(b^4)^2\right], \nonumber\\
&&v^n=-f\left[(1+a^4)b^n-b^4 a^n\right],\quad
u^n=f\left[(1+b^3)a^n-a^3 b^n\right],\\
&& \kappa=\left[(1+a^4)a^3+(1+b^3)b^4\right]/
 \left[(1+a^4)^2+(b^4)^2\right],\;\;
\chi= f\left(a^3-b^4\right).\nonumber
\end{eqnarray}

Actually eight components of ${\bf a}, {\bf b}$ depend on four real 
harmonic functions, say, $a^3, a^4, b^3, b^4$, from
which one can form two complex harmonic functions
\begin{equation}
{\cal H}_1=a^3+i(1+b^3),\quad {\cal H}_2=(1+a^4)+ib^4.
\end{equation}
Then
\begin{equation} 
f^{-1}={\rm Im}({\cal H}_1\bar {{\cal H}}_2),
\quad z=\frac{{\cal H}_1}{{\cal H}_2},\quad 
\chi=f\left({\rm Re}{\cal H}_1-{\rm Im}{\cal H}_2\right),
\end{equation} 
what gives the metric and axidilaton part of `SWIP' \cite{bko}. For
vector fields a different gauge was used in \cite{bko},
namely $v^n_\infty-iu^n_\infty =k^n$, where $k^n=k'^n+ik''^n$ is a complex
constant vector satisfying conditions $(k^n)^2=0,\, |k^n|^2=2$.
In our formalism this correspond to the following choice of the remaining
components of ${\bf a}, {\bf b}$ (consistent with 
${\bf a}^2={\bf b}^2={\bf ab}=0$):
$$a^1=k'^1a^3 + k''^1a^4,\quad 
a^2=k'^2a^3 + k''^2a^4,$$
\begin{equation}
b^1= k'^1b^3 + k''^1b^4,\quad 
b^2= k'^2b^3 + k''^2b^4\;\;,
\end{equation}
accompanied by shifts ${v'}^n=v^n+k'^n, {u'}^n=u^n-k''^n$. 
The twist potential then undergoes a transformation which makes it zero, 
while the rest of the solution will read
$$v'^n=f{\rm Re} (k^n{\cal H}_2),\quad u'^n=f{\rm Re} (k^n{\cal H}_1),$$
\begin{equation}
h_{ij}=\delta_{ij},\quad \epsilon^{ijk}\partial_j\omega_k={\rm Re}
\left[{\cal H}_1\partial_i\bar {\cal H}_2-
\bar {\cal H}_2\partial_i{\cal H}_1\right].
\end{equation}

The isotropy condition Tr$B^2=0$ in terms of charges is equivalent to the
force balance \cite{cg,gl}:
\begin{equation}\label{nf}
M^2+N^2+D^2+A^2=(Q^n)^2 + (P^n)^2.
\end{equation}
As it was noted, for $p\ge 2$ null geodesic solutions with det$B=0$
form two subclasses: those with collinear and those with non--collinear 
$\bm\alpha, \bm\beta$. In the first case $B^2=0$, hence the second condition 
arises:
\begin{equation}\label{bps2}
M^2+N^2=D^2+A^2.
\end{equation}
The conditions (53-54) together are equivalent to both BPS bounds of the 
$N=4$ theory saturated, what corresponds to the $N=2$ residual SUSY 
\cite{bko}. For non--collinear $\bm\alpha, \bm\beta$ only the force balance 
condition holds (the $N=1$ residual SUSY).

Our second example is the CPW counterpart of the DARN--NUT solution
\cite{gl}. It is well--known that certain CPW in the collision
region map onto black hole interiors \cite{chandra}. Like black holes,
CPW belong to spacetimes with two commuting Killing vectors, so
one can further specialize three-dimensional coordinates as follows
\begin{equation}
h_{ij}dx^i dx^j=e^{2\gamma} \left(d\rho^2-dz^2\right)-\rho^2 dx^2
\end{equation}
(the second Killing vector is $\partial_x$, and 
$\omega_i=\omega \delta_{ix}$ in (3)). Consider degenerate $B$, 
putting without loss of generality
$\bm\alpha^2=\bm\beta^2=1,\, \bm\alpha \bm\beta=0$ with non--collinear
$\bm\alpha,\, \bm\beta$. Then
\begin{equation}
A{\cal M}=I_4\cosh^2\sigma+
2(\bm\alpha\wedge\bm\beta)\cdot{\tilde{\bf M}}\sinh^2\sigma+
\frac{1}{2} \left(\bm\alpha\cdot\bm\Gamma^1+\bm\beta\cdot\bm\Gamma^2\right)
\sinh 2\sigma.
\end{equation}
Note that the three--space in the CPW case can not be Ricci--flat
(for the $SO(4)$ metric $\bm\alpha^2+\bm\beta^2=0$ implies
$\bm\alpha=\bm\beta=0$), with our normalization
\begin{equation}
{\cal R}_{ij}=2\sigma_i\sigma_j.
\end{equation}
Appropriate harmonic functions should be found together with $\gamma$ 
in a self--consistent way. A simple solution is
\begin{equation}
\sigma=\frac{1}{2}\ln\left(\frac{1+\tau}{1-\tau}\right),\quad 
e^{2\gamma}=\frac{1-\tau^2}{\tau^2-\zeta^2},
\end{equation}
where new coordinates correspond to
\begin{equation}
\rho^2 =(1-\tau^2)(1-\zeta^2),\quad z=-\tau\zeta.
\end{equation}
This results in the folowing family of $N=4$ CPW:
\begin{equation}
ds^2=\Sigma\left( \frac{d\tau^2}{1-\tau^2}-\frac{d\zeta^2}{1-\zeta^2}\right)-
\frac{1-\tau^2}{\Sigma}\left(dy-Q\zeta dx\right)^2 - (1-\zeta^2)\Sigma dx^2,
\end{equation}
where $\Sigma=1+(\beta^3-\alpha^4)\tau+(\alpha\wedge\beta)^{34}\tau^2$,
$Q=\beta^4+\alpha^3$ (with $\epsilon^{\tau\zeta x}=1$), 
and material fields are
$$
-v^n=\tau\left[\beta^n+\tau(\alpha\wedge\beta)^{n4}\right]/\Sigma,\quad
-u^n=\tau\left[\alpha^n+\tau(\alpha\wedge\beta)^{n3}\right]/\Sigma,$$
\begin{equation}
e^{-2\phi}=\Sigma/\Delta,\quad
\kappa=\tau\left[\beta^4-\alpha^3+
(\alpha^3\alpha^4+\beta^3\beta^4)\tau\right]/\Delta,
\end{equation}
with $\Delta=1-2\alpha^4 \tau +\left[(\alpha^4)^2+(\beta^4)^2\right]\tau^2$.
For $\alpha^1= \beta^1=\alpha^2=\beta^2=0$ this solution may be
interpreted as Ferrari--Ibanez--Bruni CPW \cite{fib} (with 
$(\alpha\wedge\beta)^{34}=1$), or as collinear impulsive CPW with dilaton
and axion ($(\alpha\wedge\beta)^{34}=-1, Q=0$). General solution (60,61)
may be considered as the CPW counterpart of the DARN--NUT solution \cite{gl}.
Indeed, if one put in the latter $r=M_0(t+1)-r_0^-,\, \cos\theta=\zeta,\, 
\varphi=x,\, t=M_0 y$  (notation of \cite{gl}), the collision region of
(60,61) is recovered with $N=-Q/(2M_0)$. Note that the extremality (BPS) 
limit of the DARN--NUT solution corresponds to $(\alpha\wedge\beta)^{34}=0$, 
the relevant CPW then has $\Sigma$ linear in $\tau$, but the above coordinate 
map becomes singular.

A notable feature of three--dimensional sigma--models on symmetric
spaces is that their further two--dimensional reduction generates
(modified) chiral equations which belong to the class of integrable
systems (for a simple derivation see, {\em e.g.} \cite{g}). Both
vacuum Einstein's and $p=1$ EMDA theory are known to admit two alternative 
Lax pairs related by the Kramer--Neugebauer (KN) map \cite{bdg}. Here we
generalize this construction to arbitrary $p$. Let indices $A, B=1,2$
correspond to coordinates on the two--surface orthogonal to Killing orbits.
Define
\begin{equation}
h_{AB}=e^{2\gamma}G_{AB}, \quad 
h_{xx}=\lambda\rho^2, \quad 
G_{AB} =
\pmatrix{
1 &0 \cr
0 & \lambda},\quad
\epsilon_{AB}=
\pmatrix{
0 &1 \cr
-1 & 0}.
\end{equation}
Introduce instead of $u^n$ the non--dualized potentials $a^n$ via
$F^n_{Ax}=\nabla_A a^n/\sqrt{2}$, and let $q^n=a^n+\omega v^n$, 
$b=B_{yx}$, (a component of the Kalb--Ramond field underlying the 
Peccei--Quinn axion $\kappa$). Then the `Matzner--Misner' counterpart of the
`potential' matrix ${\cal Q}$ for $p=2$ will be the following 
hermitean matrix
\begin{equation}
{\Omega}=
\left( \begin{array}{ccc}
\omega & -(q^1-iq^2)\\
-(q^1+iq^2)  & q^nv^n-b\\
\end{array} \right)
\end{equation}
(for $p=1$ a similar representation was given in \cite{bdg}).  Its
arbitrary--$p$ generalization is straightforward:
\begin{equation}
{\Omega}= w^0 I_k+w^a \gamma_a, \quad
2w^0 =\omega-b+q^nv^n,\quad w^n=-q^n,
\quad 2w^{p+1}=\omega+b-q^nv^n.
\end{equation}
>From the equations of motion one can derive
the following relation between ${\Omega}$ and $Q$:
\begin{equation}
\nabla Q=-\rho^{-1} \,P{\tilde \nabla} \,\Omega\,P,
\end{equation}
where $\nabla_A= (\partial_1,\partial_2),  \,{\tilde \nabla}^A
= \epsilon^{AB}\nabla_B$, and $A, B$ are raised and lowered by $G_{AB}$.
A `Matzner--Misner' matrix can now be introduced
\begin{equation}
{\cal F}=-\rho^{-1}
\left( \begin{array}{ccc}
P & -P\,\Omega\\
-\Omega \,P&\Omega \,P\, \Omega -\lambda\rho^2 P^{-1}\\
\end{array} \right),
\end{equation}
which satisfies chiral equations of the same type as ${\cal M}$:
\begin{equation}
\nabla^A \left(\rho {\cal F}^{-1}\nabla_A {\cal F}\right)=0,\quad
\nabla^A \left(\rho {\cal M}^{-1}\nabla_A {\cal M}\right)=0.
\end{equation}

Variables entering ${\cal F}$ are related nonlocally  
to the sigma--model variables in ${\cal M}$. Now, by definition,
a KN map is a {\em local} relation between two alternative
forms of chiral equations.
Comparing (13) and (65) one finds that the map
\begin{equation}
{\cal Q}\rightarrow \sqrt{-\lambda}\Omega,
\quad {\cal P}\rightarrow \rho{\cal P}^{-1},
\end{equation}
transform the equations for $({\cal P,\, Q})$ into those for
$({\cal P},\, \Omega)$. This opens a way of further development as
discussed in \cite{bdg}.

Hence the Ernst--type picture of the $N=4$ supergravity amounts to the 
{\it pseudounitary} embedding of the three--dimensional U--duality group.
Previously found symplectic representation of the EMDA theory
is valid uniquely for the one--vector truncation. Meanwhile its basic
features such as an existence in the two--dimensional case of the 
Matzner--Misner counterpart and the Kramer--Neugebauer mapping remain valid 
thus opening the way to application of various techniques of the theory
of integrable systems.

\bigskip
One of the authors (DVG) is grateful to the Theory Division, CERN for
hospitality, while the work was in progress. Stimulating discussions
with I. Bakas are acknowledged. DVG thanks G. Clement for clarifying
detailes of the geodesic technique, and R. Kallosh for helpful 
correspondence. The work was supported by the RFBR Grant 96--02--18899.


\bigskip
\bigskip
\begin{thebibliography}{99}
\bibitem{tudo}
G.W. Gibbons, Nucl.Phys. {\bf B204}, 337 (1982);
G.W. Gibbons and K. Maeda, Nucl. Phys. {\bf B298}, 741 (1988);
D. Garfinkle, G.T. Horowitz, and A. Strominger Phys. Rev. {\bf D43}, 3140
(1991); {\bf D45}, 3888 ({\bf E}) (1992);
R. Kallosh, A. Linde, T. Ortin, A. Peet, and A. van Proyen,
Phys. Rev. {\bf D 46}, 5278 (1992);
R. Kallosh, D. Kastor, T. Ortin, and T. Torma,
Phys. Rev. {\bf D 50}, 6374 (1994);
M. Cveti{\v c} and A. Tseytlin, Phys. Rev. {\bf D 53}, 5619 (1996),
`{\it Non--Extreme Black Holes from Non--Extreme Intersecting M--branes}',
DAMTP/R/96/27, hep-th/9606033.
\bibitem{ad}
D. Gal'tsov, '{\it Square of General Relativity}' , Proc. of the First
Australasian Conf. on Gen. Rel. and Grav., Adelaide, February 12--17 1996,
hep-th/9608021.
\bibitem{gk}
D.V. Gal'tsov and O.V. Kechkin, Phys. Rev. {\bf D50}, 7394 (1994);
Phys. Lett. {\bf B361}, 52 (1995).
\bibitem{g}
D.V. Gal'tsov, Phys. Rev. Lett. {\bf 74}, 2863 (1995)
(hep-th/9410217).
\bibitem{eh}
J. Ehlers, in {\em Les Theories Relativistes de la
Gravitation}, CNRS, Paris, 1959, p. 275.
\bibitem{gpr}
A. Giveon, M. Porrati, and E. Rabinovici, Phys. Rept.
{\bf 244}, 77 (1994) .
\bibitem{udu}
D.V. Gal'tsov, {\em `Symmetries of Heterotic String Effective Theory
in Three and Two Dimensions'}, in  
`{\sl Heat Kernel Techniques and Quantum Gravity}', (Int. Workshop, Winnipeg,
Canada, 2---6 August, 1994),  S.~A. Fulling (ed.),
Discourses in Mathematics and Its Applications,
No.~4, Texas A\&M University,
College Station, Texas, 1995, pp.~423--449 (hep-th/9606042);
D.V. Gal'tsov and O.V. Kechkin,
Phys. Rev. D. {\bf D54}, 1656 (1996).
\bibitem{ht}
C. Hull and P. Townsend, Nucl. Phys. {\bf B450}, 109 (1995).
\bibitem{cg}
G. Clement and D. Gal'tsov,
Phys. Rev. D. {\bf D54}, 6136 (1996).
\bibitem{gl}
D.V. Gal'tsov and P.S. Letelier, 
`{\it Reissner--Nordstr\"om type black holes in dilaton--axion gravity}',
gr-qc/9608023;
`{\it Ehlers--Harrison transformations and 
black holes in dilaton--axion gravity with multiple vector fields},
gr-qc/9612007.
\bibitem{bko}
E. Bergshoeff, R. Kallosh, and T. Ortin,
Nucl. Phys. {\bf B478}, 165 (1996).
\bibitem{chandra}
S. Chandrasekhar and B.C. Xanthopoulos,
Proc. Roy. Soc. Lond. {\bf A410}, 311 (1987).
\bibitem{fib}
V. Ferrari, J. Ibanez, and M. Bruni,
Phys. Rev. D. {\bf D36}, 1053 (1987).
\bibitem{bdg}
D.V. Gal'tsov, {\it Geroch--Kinnersley--Chitre group for
Dilaton--Axion Gravity}, in `{\sl Quantum Field Theory under the Influence of
External Conditions}', M. Bordag (Ed.) (Proc. of the International Workshop,
Leipzig, Germany, 18--22 September 1995), B.G. Teubner Verlagsgessellschaft,
Stuttgart--Leipzig, 1996, pp. 228-237, (hep-th/9606041).

\end{thebibliography}
\end{document}